\documentclass[12pt]{article}

\usepackage{epsfig}

\usepackage{bbm}
\usepackage{amssymb,amsmath}

\newcommand{\be}{\begin{equation}}
\newcommand{\ee}{\end{equation}}
\newcommand{\ba}{\begin{eqnarray}}
\newcommand{\ea}{\end{eqnarray}}
\newcommand{\no}{\nonumber\\}

\newcommand{\mnu}{\mathcal{M}_\nu}

\newcommand{\mew}{m_\mathrm{ew}}

\begin{document}

\title{\normalsize \hfill UWThPh-2009-1 \\[8mm]
\LARGE Type~II seesaw mechanism for Higgs doublets \\
and the scale of new physics}

\author{
W.~Grimus,$^{(1)}$\thanks{E-mail: walter.grimus@univie.ac.at}
\
L.~Lavoura$\, ^{(2)}$\thanks{E-mail: balio@cftp.ist.utl.pt}
\ and
B.~Radov\v{c}i\'c$\, ^{(1,3)}$\thanks{E-mail: bradov@phy.hr}
\\*[3mm]
\small $^{(1)}$ University of Vienna, Faculty of Physics \\
\small Boltzmanngasse 5, A--1090 Vienna, Austria
\\*[2mm]
\small $^{(2)}$ Technical University of Lisbon \\
\small Centre for Theoretical Particle Physics, 1049-001 Lisbon, Portugal
\\*[2mm]
\small $^{(3)}$ University of Zagreb, Faculty of Science,
Department of Physics \\
\small P.O.B.~331, HR--10002 Zagreb, Croatia
}

\date{10 March 2009}

\maketitle

\begin{abstract}
We elaborate on an earlier proposal by Ernest Ma
of a type~II seesaw mechanism
for suppressing the vacuum expectation values of some Higgs doublets.
We emphasize that,
by nesting this form of seesaw mechanism into various other seesaw mechanisms,
one may obtain light neutrino masses in such a way that
the new-physics scale present in the seesaw
mechanism---the masses of
scalar gauge-$SU(2)$ triplets,
scalar $SU(2)$ doublets,
or right-handed
neutrinos---does not need
to be higher than a few $10 \, \mathrm{TeV}$.
We also investigate other usages
of the type~II seesaw mechanism for Higgs doublets.
For instance,
the suppression of the vacuum expectation values of Higgs doublets may
realize Froggatt--Nielsen suppression factors
in some entries of the fermion mass matrices.
\end{abstract}

\newpage

\section{Introduction}

The type~I seesaw mechanism~\cite{seesaw}
is a favourite with high-energy physicists
for explaining why the neutrino masses are so tiny.
Unfortunately,
in the usual realization of that mechanism
the scale $m_R$ of the Majorana masses
of the right-handed neutrinos $\nu_R$ should be
$10^{13} \,\mbox{GeV}$ (assuming that
the natural scale of the
neutrino Dirac mass matrix is the electroweak scale).
As a consequence,
the possibility of direct tests of the seesaw mechanism seems very remote.
Lowering $m_R$ to the TeV scale,
although desirable from the point of view
of experimental tests of the type~I seesaw mechanism,
apparently contradicts the aim with which it was invented,
since it would require artificially suppressing
the Yukawa couplings of the $\nu_R$ to values of order $10^{-5}$.

Another mechanism for explaining the smallness of the neutrino masses
is the type~II seesaw mechanism~\cite{II},
which suppresses the vacuum expectation values (VEVs)
of the neutral components of scalar gauge-$SU(2)$ triplets,
in such a way that the left-handed neutrinos $\nu_L$,
which acquire Majorana masses
from their Yukawa couplings to those neutral components,
are extremely light.
Just like the type~I seesaw mechanism,
the type~II seesaw mechanism requires
a very high mass scale,
which now occurs in the mass terms of the scalar triplets.
Those large mass terms make the scalar triplets extremely heavy
and therefore the type~II seesaw mechanism,
like the type~I seesaw mechanism,
is very difficult to test experimentally.

In the general case,
for instance in Grand  Unified Theories based on the gauge group $SO(10)$,
both type~I and type~II seesaw mechanisms are present~\cite{generalD+M}.

Several proposals have been made
to bring the high mass scale of the seesaw mechanism(s) down to the TeV range,
so that they might be experimentally testable,
for instance at the Large Hadron Collider at CERN.
The most straightforward possibility
is to have cancellations within the type~I seesaw mechanism
such that $m_R$ may be relatively low
without the need to excessively suppress
the Yukawa couplings~\cite{cancellation within I};
the general conditions for this to happen were given in~\cite{smirnov}.
Cancellations between the type~I and type~II seesaw
contributions to the neutrino masses
have also been considered~\cite{cancellationI+II}.
In the ``inverse seesaw mechanism''~\cite{inverse-seesaw}
there is both a high scale in the TeV range and a low scale in the keV range.
Other proposals include radiative neutrino masses
generated by three-loop diagrams~\cite{3-loop}
or a specific type of mirror fermions~\cite{mirror-fermions}.

In this letter we develop a proposal originally made in~\cite{ma1}.
We elaborate on its two separate ideas:
\begin{enumerate}
\renewcommand{\labelenumi}{\roman{enumi}.}
\item A type~II seesaw mechanism suppresses the VEVs of some
\emph{Higgs doublets}.
\item A nesting of that type~II seesaw mechanism
\emph{inside some other seesaw mechanism}
(which may be of any type)
allows one to lower the high mass scale of that seesaw mechanism.
\end{enumerate}
The aim of this letter is to generalize the proposal 
of~\cite{ma1} in several directions:
\begin{itemize}
\item We describe (in section 2) the \emph{general mechanism}
for suppressing Higgs-doublet VEVs and give several examples thereof.
\item We show (in section 3)
that this ``type~II seesaw mechanism for Higgs doublets''
may be nested \emph{inside various seesaw mechanisms}.\footnote{The
original proposal~\cite{ma1}
was a type~II seesaw mechanism for Higgs doublets
within a type~I seesaw mechanism.
A later suggestion~\cite{ma2}
was a type~II seesaw mechanism for Higgs doublets
within a type~III seesaw mechanism.}
We propose in particular a type~II seesaw mechanism for Higgs doublets
inside the usual type~II seesaw mechanism for scalar triplets.
\item We develop (in section 4)
a multiply nested type~II seesaw mechanism for many Higgs doublets
which may mimic the Froggatt--Nielsen mechanism~\cite{FN}.
This suggests new ways of explaining
the relative smallness of some charged-fermion masses---without the need
for new heavy fermions
as in the seesaw mechanism for Dirac fermions~\cite{dirac-seesaw}.
\end{itemize}
In summary,
the message that we want to convey in this letter is that,
by using the nesting of seesaw mechanisms,
a heavy mass scale $m_H$
many orders of magnitude larger than
the electroweak scale $\mew \sim 100 \, \mathrm{GeV}$
is not compelling;
an $m_H$ just two or three orders of magnitude above $\mew$ may suffice.

\section{Type II seesaw mechanism for Higgs doublets}
\label{HDII}

Consider a model with several Higgs doublets
$\phi_j = ( \phi_j^+, \ \phi_j^0 )^T$,
$j = 1, \ldots, n_h$.
The VEVs of the Higgs doublets are of the form
\be
\label{vevs}
\left\langle \phi_j \right\rangle_0
= \left( \begin{array}{c} 0 \\ v_j \end{array} \right).
\ee
We assume that $\left| v_1 \right| \sim \mew$.
Our aim is to produce a seesaw mechanism to suppress $\left| v_2 \right|$.
We write the scalar potential as
\be
\label{V}
V = \sum_{j=1}^{n_h} \mu_j^2 \phi_j^\dagger \phi_j
+ \left( V_l + V_l^\dagger \right) + V_r.
\ee
We assume that $\mu_1^2 < 0$ and that $\left| \mu_1^2 \right| \sim \mew^2$
in order to generate a spontaneous symmetry breaking
leading to $\left| v_1 \right| \sim \mew$.
On the other hand,
we assume that $\mu_2^2 > 0$ and that $\mu_2^2 = m_H^2 \gg \mew^2$.
In equation~(\ref{V}) $V_l$ represents some terms linear in $\phi_2$
which we assume to be present in $V$.
All the remainder of $V$,
\textit{i.e.}\ everything but the mass terms for the Higgs doublets
and the terms $V_l$ and $V_l^\dagger$ linear in $\phi_2$ and $\phi_2^\dagger$,
respectively,
is denoted $V_r$;
in the simplest cases $V_r$ will consist only of quartic terms.

Inserting the VEVs into the potential one has
\be
\left\langle V \right\rangle_0
= \sum_{j=1}^{n_h} \mu_j^2 \left| v_j \right|^2
+ A v_2 + A^\ast v_2^\ast + \left\langle V_r \right\rangle_0,
\ee
where $A$ has the dimension of the cube of a mass.
Then,
despite the positiveness of $\mu_2^2$,
a non-vanishing VEV $v_2$ is induced,
approximately given by
\be
v_2 \approx - \frac{A^\ast}{\mu_2^2}.
\ee
The quantity $A$ depends on the specific model.
It has to contain at least one $v_j \neq v_2$
and this $v_j$ will in general be of order $m_\mathrm{ew}$.
If we assume that $\mu_2^2$ is
the only parameter in the scalar potential of order $m_H^2$,
then we expect $|A| \sim m_\mathrm{ew}^3$.
In this case $\left| v_2 \right| \sim m_\mathrm{ew}^3 / m_H^2$
is suppressed by \emph{two} powers of $\mew$ over $m_H$,
where $m_H$ is the scale of new physics.

\paragraph{Two Higgs doublets and a softly broken symmetry:}
In the original proposal~\cite{ma1}
of the type~II seesaw mechanism for Higgs doublets
there were only two Higgs doublets and
no other scalar multiplets.\footnote{A related scenario
with the assumption $\mu_1^2 = 0$
was proposed in~\cite{calmet}.}
A $U(1)$ symmetry
\be
\label{u1}
\phi_2 \to e^{i \alpha} \phi_2
\ee
was softly broken in the scalar potential by
\be
\label{Va}
V_l = \mu^2 \phi_1^\dagger \phi_2.
\ee
Then,
\be\label{v2soft}
v_2 \approx - \frac{{\mu^2}^\ast v_1}{\mu_2^2}.
\ee
The VEV $v_1$ alone must produce the $W^\pm$ and $Z^0$ masses,
therefore $\left| v_1 \right| \approx 174 \, \mathrm{GeV} \sim \mew$.
We assume that $\mu_2^2 = m_H^2 \gg \mew^2$
and that $\left| \mu^2 \right| \lesssim m_H^2$,
where the symbol $\lesssim$ means ``not much larger than''.
We may assume that $\left| \mu^2 \right| \sim \mew^2$
and then $v_2$ is suppressed by two powers of $\mew / m_H$
relative to $v_1$.\footnote{As a matter of fact,
since $V_l$ in this case breaks softly the symmetry~(\ref{u1}),
it would be technically natural to assume $\left| \mu^2 \right| \ll \mew^2$,
as was done in~\cite{ma1},
and then $\left| v_2 \right|$ would be even smaller.}

\paragraph{General two-Higgs-doublet model:}
Actually,
one could dispense with any symmetry
and consider the general two-Higgs-doublet model,
employing the same assumptions as in the previous paragraph.
Then in $V_l$ not only the term of equation~(\ref{Va}) is present but also
\be
\left( \phi_1^\dagger \phi_1 \right) \left( \phi_1^\dagger \phi_2 \right).
\ee
Therefore,
one has two sources which induce a non-zero $v_2$.
As discussed in~\cite{trott},
one obtains a suppression factor of $v_2$
of the same order of magnitude as before.

\paragraph{Two Higgs doublets and a scalar singlet:}
If we dislike soft symmetry breaking
the simplest alternative is to introduce into the theory
a complex scalar gauge singlet $\chi$ with VEV $v_\chi$.
The $U(1)$ symmetry~(\ref{u1}) becomes
\be
\label{u1'}
\phi_2 \to e^{i \alpha} \phi_2,
\quad
\chi \to e^{- i \alpha} \chi.
\ee
Then,
\ba
V_l &=& m \phi_1^\dagger \phi_2 \chi,
\\
v_2 &\approx& - \frac{m^\ast
v_\chi^\ast v_1}
{\mu_2^2}.
\ea
There is a large degree of arbitrariness
in the orders of magnitude of $\left| m \right|$
and of the VEV of $\chi$,
but we may conservatively assume them to be of order $\mew$.
Then once again $\left| v_2 / v_1 \right| \sim \left( \mew / m_H \right)^2$.

\paragraph{Symmetry $\mathbbm{Z}_2$ instead of $U(1)$:}
Instead of the $U(1)$ symmetry~(\ref{u1}) originally used in~\cite{ma1}
one may employ the weaker symmetry
\be
\label{z2}
\phi_2 \to - \phi_2, \quad \chi \to - \chi.
\ee
In this case $\chi$ may as a matter of fact be a real field.
The symmetry~(\ref{z2}) allows for a richer scalar potential,
with extra terms $( \phi_1^\dagger \phi_2 )^2$
and $\chi^4$ and their Hermitian conjugates.

\paragraph{Three Higgs doublets:}
A more complicated model has three Higgs doublets and a symmetry.
\be
\label{z4}
\mathbbm{Z}_4: \quad \phi_2 \to -\phi_2, \quad \phi_3 \to i \phi_3.
\ee
Note that we now assume $\mu_j^2 < 0$ and
$\left| \mu_j^2 \right| \sim \mew^2$ for both $j = 1,3$.
Then
\ba
V_l &=& \lambda \left( \phi_3^\dagger \phi_2 \right)
\left( \phi_3^\dagger \phi_1 \right),
\\
V_r &=& \lambda^\prime \left( \phi_1^\dagger \phi_2 \right)^2 + \cdots,
\\
v_2 &\approx& - \frac{\lambda^\ast v_1^\ast v_3^2}{\mu_2^2}.
\ea
In this case $\left| v_1 \right|$ and $\left| v_3 \right|$
are \emph{necessarily} of order $\mew$
(or smaller)
and one needs no extra assumption to conclude that
$\left| v_2 \right| \sim \mew^3 / m_H^2$.

Before we proceed to investigate the nesting of seesaw mechanisms,
we want to mention some simple applications
of a seesaw mechanism for Higgs doublets.
Suppose that
$\phi_2$ has Yukawa couplings only to the $\nu_R$,
and $\phi_1$ to all charged fermionic gauge-$SU(2)$ singlets.
Then with the small VEV $v_2$ we have the option of
a seesaw mechanism for \emph{Dirac} neutrinos,
if we dispense with a $\nu_R$ Majorana mass term.
We would then use
$\left| v_2 \right| \sim \mew^3 / m_H^2 \sim 1\, \mbox{eV}$,
where we assume
$1 \, \mathrm{eV}$ to be the scale of the light-neutrino masses,
obtaining the estimate $m_H \sim \sqrt{10^{33}} \, \mathrm{eV} =
10^{7.5} \,\mathrm{GeV}$.
We could also try to ``explain'' the smallness
of the down-type-quark masses as compared to the up-type-quark masses
by enforcing the coupling of $\phi_1$ to the up-quark singlets
and $\phi_2$ to the down-quark singlets in the Yukawa couplings.
Assuming
\be
\left| \frac{v_2}{v_1} \right| \sim \frac{m_b}{m_t} \sim
\left( \frac{\mew}{m_H} \right)^2,
\ee
we find for the mass of the heavy Higgs doublet
$m_H \sim 6 \, \mew$.

\section{Nesting of seesaw mechanisms}

\subsection{Type~I seesaw mechanism}

The type~I seesaw formula is~\cite{seesaw}
\be
\label{mnu}
\mnu = - M_D^T M_R^{-1} M_D,
\ee
where $\mnu$ is the effective $\nu_L$ Majorana mass matrix,
$M_R$ is the Majorana mass matrix of the $\nu_R$
and $M_D$ is the Dirac mass matrix connecting the $\nu_R$ to the $\nu_L$.
This Dirac mass matrix is generated by Yukawa couplings
\be
\label{yuk}
\mathcal{L}_\mathrm{Yukawa} =
\bar \nu_R \tilde \phi_2^\dagger Y D_L
+ \mathrm{H.c.},
\ee
where $Y$ is a matrix (in flavour space) of Yukawa coupling constants,
$D_L = ( \nu_L, \ell_L )^T$ are $SU(2)$ doublets of left-handed leptons,
$\phi_2$ is the Higgs doublet
whose VEV is suppressed by a type~II seesaw mechanism
and $\tilde \phi_2 \equiv i \tau_2 \phi_2^\ast$.

In order for the Yukawa couplings of the $\nu_R$ in equation~(\ref{yuk})
to involve only the Higgs doublet $\phi_2$,
one needs to suitably extend the symmetries $U(1)$,
$\mathbbm{Z}_2$ or $\mathbbm{Z}_4$ of the previous section.
In the case of the $U(1)$ symmetry,
one must add $D_L \to e^{- i \alpha} D_L$
and $\ell_R \to e^{- i \alpha} \ell_R$ to the assignment~(\ref{u1'})
(the $\ell_R$ are the right-handed charged-lepton singlets)~\cite{ma1}.
In the case of the $\mathbbm{Z}_2$ or $\mathbbm{Z}_4$ symmetries,
one must add $\nu_R \to - \nu_R$ to the assignments~(\ref{z2}) and~(\ref{z4}),
respectively.

It follows from equation~(\ref{yuk}) that $M_D = v_2 Y$,
hence $\mnu = - v_2^2\, Y^T M_R^{-1} Y$.
As before,
we assume the matrix elements of $\mnu$ to be of order eV.
If we allowed the VEV $v_2$
to be of order the electroweak scale $\mew \sim 100 \, \mathrm{GeV}$,
and assuming the Yukawa coupling constants to be of order unity,
we would find the scale $m_R$ of $M_R$
to be of order $10^{13} \, \mathrm{GeV}$,
as advertised in the introduction.
Lowering $m_R$ to the TeV scale while keeping $v_2 \sim \mew$ requires
(assuming no cancellation mechanism)
the Yukawa couplings to be of order $10^{-5}$,
as also advertised in the introduction.
But if $v_2 \sim \mew^3 / m_H^2$ is suppressed
by a type~II seesaw mechanism for Higgs doublets,
as first proposed in~\cite{ma1},
then $1 \, \mathrm{eV} \sim \mew^6 / \left( m_R m_H^4 \right)$
even with Yukawa coupling constants of order unity.
This represents a \emph{fivefold suppression} of the neutrino masses.
Assuming for simplicity $m_R = m_H$,
one obtains
\be
m_H \sim \sqrt[5]{10^{66}}\, \mathrm{eV} \approx 16\, \mathrm{TeV}.
\ee

\subsection{Type~II seesaw mechanism}

In the type~II seesaw mechanism~\cite{II}, a scalar gauge-$SU(2)$ triplet
\begin{equation}
\Delta = \left( \begin{array} {cc}
 \delta^+/\sqrt{2}  &  \delta^{++} \\
 \delta^0 & - \delta^+/\sqrt{2}
\end{array} \right)
\end{equation}
is introduced such that the $\nu_L$ acquire Majorana masses through
the
VEV of the neutral component of the scalar triplet:
\be
\left\langle \Delta \right\rangle_0 = \left( \begin{array} {cc}
 0  &  0 \\
 v_\Delta & 0
\end{array} \right).
\ee
This VEV is induced by the term linear in $\Delta$ in the scalar
potential and is suppressed by the high mass of the scalar triplet.

In order for the terms linear in $\Delta$ to involve only the Higgs
doublet $\phi_2$ whose VEV is suppressed
by the type~II seesaw mechanism discussed in section~\ref{HDII},
we introduce a $\mathbbm{Z}_4$ symmetry\footnote{In~\cite{ma3}
a softly broken $U(1)$ symmetry has been used instead,
together with assumptions
on the soft-breaking parameters in the scalar potential.}:
\be
\label{z4II}
\phi_2 \to i  \phi_2,
\quad
\Delta \to - \Delta.
\ee
We write the scalar potential as
\be
\label{VII}
V =
\sum_{j=1}^2 \mu_j^2 \phi_j^\dagger \phi_j +
\mu_\Delta^2 \mbox{Tr} \left( \Delta^\dagger\Delta \right) +
\left( \mu^2\phi_1^\dagger \phi_2 + \mbox{H.c.} \right) +
\left( \mu' \phi_2^\dagger \Delta \tilde \phi_2 + \mbox{H.c.} \right) + V_q,
\ee
where $V_q$ consists only of quartic terms. The $\mathbbm{Z}_4$
symmetry is softly broken by operators of dimension two. Instead of
a softly broken symmetry for the type~II seesaw mechanism for the VEV
of $\phi_2$, one could employ one of the alternatives given in
section~\ref{HDII}. In order to have a Dirac mass term for charged
leptons and a Majorana
mass term for $\nu_L$ generated by VEVs of $\phi_1$ and $\Delta$,
respectively, one needs to extend the $\mathbbm{Z}_4$ symmetry in
equation~(\ref{z4II}) to $D_L \to i D_L$
and $\ell_R \to i \ell_R$.

Now we proceed according to section~\ref{HDII}. On the one hand,
we assume that $\mu_1^2 < 0$ and that $\left| \mu_1^2 \right| \sim \mew^2$
in order to generate a spontaneous symmetry breaking with
$\left| v_1 \right| \sim \mew$.
On the other hand, we require
\be\label{condd}
\mu_2^2 > 0, \quad \mu_\Delta^2 > 0
\quad \mbox{and} \quad
\mu_2^2 \sim \mu_\Delta^2 \sim  m_H^2 \gg \mew^2.
\ee
The terms linear in $\phi_2$ and $\Delta$ in equation~(\ref{VII})
generate non-vanishing VEVs $v_2$ and $v_\Delta$, respectively.
Using the result for $v_2$ of equation~(\ref{v2soft}), the VEV of
$\Delta$ is given by
\be
v_\Delta \approx - \frac{{\mu'}^\ast v_2^2}{\mu_\Delta^2} \approx -
\frac{{\mu'}^\ast \left({\mu^2}^\ast\right)^2 v_1^2}{\mu_2^4\, \mu_\Delta^2}.
\ee
As before, there is a degree of arbitrariness
in the orders of magnitude of
$\left| \mu \right|$ and $\left| \mu' \right|$,
but we may assume them to be of order $\mew$. Then
$v_\Delta$ is  suppressed by six powers of $\mew / m_H$
relative to $v_1$. Keeping the Yukawa coupling constants of order
unity, this represents a \emph{sixfold suppression} of the neutrino
masses. Assuming again the matrix elements of $\mnu$ to be of order
eV which amounts to $v_\Delta \sim 1\,\mbox{eV}$,
with equation~(\ref{condd}) we estimate
\be
m_H \sim \sqrt[6]{10^{77}}\, \mathrm{eV} \approx 7\, \mathrm{TeV}.
\ee
By raising the mass of the $\phi_2$ to $20\, \mathrm{TeV}$,
one shifts $\mu_\Delta$ below $1\, \mathrm{TeV}$,
and the $\delta^{++}$,
whose mass is just $\mu_\Delta$,
could possibly be within reach of the LHC---see for instance~\cite{schwetz}
and the references therein.

\section{Multiple nesting of type~II seesaw mechanisms
for Higgs doublets}

In this section we show that a multiple nesting
of successive type~II seesaw mechanisms for several Higgs doublets
is able to realize Froggatt--Nielsen~\cite{FN} suppression factors
by using only Higgs doublets and renormalizable interactions.

As an example,
we consider the hierarchy of 
charged-fermion masses:
\be
\begin{array}{rcl}
m_t &\sim& \mew
\\
m_b, m_c, m_\tau &\sim& 2 \, \mathrm{GeV},
\\
m_s, m_\mu &\sim& 0.1 \, \mathrm{GeV},
\\
m_u, m_d &\sim& 0.005 \, \mathrm{GeV},
\\
m_e &\sim& 0.0005 \, \mathrm{GeV}.
\end{array}
\ee
This hierarchy suggests that the charged-fermion mass matrices
may involve a suppression factor $\epsilon \sim 1/20$
according to the pattern
\be
M_u \sim \left( \begin{array}{ccc}
1 & \epsilon & \epsilon^3 \\
1 & \epsilon & \epsilon^3 \\
1 & \epsilon & \epsilon^3
\end{array} \right),
\quad
M_d \sim \left( \begin{array}{ccc}
\epsilon & \epsilon^2 & \epsilon^3 \\
\epsilon & \epsilon^2 & \epsilon^3 \\
\epsilon & \epsilon^2 & \epsilon^3
\end{array} \right),
\quad
M_\ell \sim \left( \begin{array}{ccc}
\epsilon & \epsilon^2 & \epsilon^4 \\
\epsilon & \epsilon^2 & \epsilon^4 \\
\epsilon & \epsilon^2 & \epsilon^4
\end{array} \right).
\ee
The suppression factors in the various elements
of these mass matrices may be explained
\textit{\`a la} Froggatt--Nielsen~\cite{FN}
as the result of a spontaneously broken horizontal symmetry.
We suggest to view them instead as the product of
a nested type~II seesaw mechanism for Higgs doublets.\footnote{A
  similar idea was already put forward in~\cite{ma5} and subsequently 
  combined with the leptonic model of~\cite{ma1} in a  
  supersymmetric way~\cite{ma4}.}

We postulate the existence of six Higgs doublets $\phi_{1, \ldots, 6}$,
where $\phi_1$ and $\phi_2$ have VEVs of order $\mew$
and Yukawa couplings which generate the first column of $M_u$,
$\phi_3$ has VEV of order $\epsilon \mew$ and generates
the second column of $M_u$ and the first columns of $M_d$ and $M_\ell$,
$\phi_4$ has VEV of order $\epsilon^2 \mew$ and its Yukawa couplings
yield the second columns of $M_d$ and $M_\ell$,
and so on.

We implement the hierarchy of VEVs in the following way.
The scalar potential is of the form
\be
V =
\sum_{j=1}^6
\left( \mu_j^2 + \frac{\lambda_j}{2} \phi_j^\dagger \phi_j \right)
\phi_j^\dagger \phi_j
+ \sum_{j<k} \left(
\lambda^{\prime}_{jk} \phi_j^\dagger \phi_j \phi_k^\dagger \phi_k
+
\lambda^{\prime \prime}_{jk} \phi_j^\dagger \phi_k \phi_k^\dagger \phi_j
\right)
+ V_t + V_t^\dagger.
\ee
We assume that $\mu_1^2$ and $\mu_2^2$ are both negative
and of order $\mew^2$,
while $\mu_{3, \ldots, 6}^2$ are positive and of order $m_H^2$,
with $\left( \mew / m_H \right)^2 \sim \epsilon$.\footnote{With
$\epsilon \sim 1/20$ this produces only a slight difference
between $\mew$ and $m_H$.
This certainly constitutes a drawback of the present model.}
The VEV of $\phi_3$ is induced out of the VEVs of $\phi_1$ and $\phi_2$
via a term
\be
\label{kappa1}
\kappa_1 \phi_1^\dagger \phi_2 \phi_1^\dagger \phi_3
\ee
in $V_t$.
This leads to $v_3 \approx - \kappa_1^\ast v_1^2 v_2^\ast / \mu_3^2$.
Since the coupling constant $\left| \kappa_1 \right| \lesssim 1$,
$\left| v_3 \right|$ is of order $\mew^3 / m_H^2$.
Afterwards the VEV of $\phi_4$ is induced by a further term in $V_t$,
\be
\label{kappa2}
\kappa_2 \phi_2^\dagger \phi_3 \phi_2^\dagger \phi_4.
\ee
This leads to $v_4 \approx - \kappa_2^\ast v_2^2 v_3^\ast / \mu_4^2
\sim \mew^5 / m_H^4$.
The VEVs $v_5$ and $v_6$ are successively induced by terms
\ba
& & \kappa_3 \phi_1^\dagger \phi_4 \phi_1^\dagger \phi_5,
\label{kappa3}
\\
& & \kappa_4 \phi_2^\dagger \phi_5 \phi_2^\dagger \phi_6,
\label{kappa4}
\ea
respectively,
in $V_t$.\footnote{Instead of the terms~(\ref{kappa1})--(\ref{kappa4})
we might imagine other possibilities.
The present text thus constitutes only a proof of the viability
of the mechanism.}

In order to make sure that there are in $V_t$ no other terms
which might induce larger (unsuppressed) VEVs,
we must impose a symmetry $\mathcal{S}$ on the theory.
For simplicity we assume that symmetry to be Abelian:
\be
\mathcal{S}: \quad \phi_j \to \sigma_j \phi_j,
\ee
with $\left| \sigma_j \right| = 1$ for $j = 1, \ldots, 6$.
We assume,
of course,
the six factors $\sigma_{1,  \ldots, 6}$ to be all different.
In order for the four terms~(\ref{kappa1})--(\ref{kappa4}) to be allowed,
we must assume
\be
\sigma_1^2 = \sigma_2 \sigma_3, \quad
\sigma_2^2 = \sigma_3 \sigma_4, \quad
\sigma_1^2 = \sigma_4 \sigma_5, \quad
\sigma_2^2 = \sigma_5 \sigma_6.
\ee
Therefore,
\be
\sigma_3 = \frac{\sigma_1^2}{\sigma_2}, \quad
\sigma_4 = \frac{\sigma_2^3}{\sigma_1^2}, \quad
\sigma_5 = \frac{\sigma_1^4}{\sigma_2^3}, \quad
\sigma_6 = \frac{\sigma_2^5}{\sigma_1^4}.
\ee
It follows that the bilinears $\phi_j^\dagger \phi_k$ ($j < k$)
transform as
\be
\begin{array}{rlcrlrclrclrcl}
\phi_1^\dagger \phi_2: & {\displaystyle \frac{\sigma_2}{\sigma_1}}, &
 & & & & & & & & & & &
\\*[3mm]
\phi_1^\dagger \phi_3: & {\displaystyle \frac{\sigma_1}{\sigma_2}}, &
 & \phi_2^\dagger \phi_3: & {\displaystyle \frac{\sigma_1^2}{\sigma_2^2}}, &
 & & & & & & & &
\\*[3mm]
\phi_1^\dagger \phi_4: & {\displaystyle \frac{\sigma_2^3}{\sigma_1^3}}, &
 & \phi_2^\dagger \phi_4: & {\displaystyle \frac{\sigma_2^2}{\sigma_1^2}}, &
 & \phi_3^\dagger \phi_4: & {\displaystyle \frac{\sigma_2^4}{\sigma_1^4}}, &
 & & & & &
\\*[3mm]
\phi_1^\dagger \phi_5: & {\displaystyle \frac{\sigma_1^3}{\sigma_2^3}}, &
 & \phi_2^\dagger \phi_5: & {\displaystyle \frac{\sigma_1^4}{\sigma_2^4}}, &
 & \phi_3^\dagger \phi_5: & {\displaystyle \frac{\sigma_1^2}{\sigma_2^2}}, &
 & \phi_4^\dagger \phi_5: & {\displaystyle \frac{\sigma_1^6}{\sigma_2^6}}, &
 & &
\\*[3mm]
\phi_1^\dagger \phi_6: & {\displaystyle \frac{\sigma_2^5}{\sigma_1^5}}, &
 & \phi_2^\dagger \phi_6: & {\displaystyle \frac{\sigma_2^4}{\sigma_1^4}}, &
 & \phi_3^\dagger \phi_6: & {\displaystyle \frac{\sigma_2^6}{\sigma_1^6}}, &
 & \phi_4^\dagger \phi_6: & {\displaystyle \frac{\sigma_2^2}{\sigma_1^2}}, &
 & \phi_5^\dagger \phi_6: & {\displaystyle \frac{\sigma_2^8}{\sigma_1^8}}.
\end{array}
\label{listing}
\ee
We assume that all the factors in this list are different
from unity---else there would be (at least) two Higgs doublets
transforming identically under $\mathcal{S}$---and also different
from each other---so that there are as few terms as possible in $V_t$.
This requires
\be
\sigma_1^p \neq \sigma_2^p \ \mbox{for} \ p = 1, 2, \ldots, 14.
\ee
Therefore we must choose for $\mathcal{S}$ a group $\mathbbm{Z}_n$
with $n > 14$.
It is enough to choose $\mathcal{S} = \mathbbm{Z}_{15}$ with
\be
\phi_1 \to \omega \phi_1, \
\phi_2 \to \omega^2 \phi_2, \
\phi_3 \to \phi_3, \
\phi_4 \to \omega^4 \phi_4, \
\phi_5 \to \omega^{13} \phi_5, \
\phi_6 \to \omega^6 \phi_6,
\ee
where $\omega \equiv \exp{\left( 2 i \pi / 15 \right)}$.
From the list~(\ref{listing}) we learn that the full $V_t$ is
\ba
V_t &=&
\kappa_1 \phi_1^\dagger \phi_2 \phi_1^\dagger \phi_3
+ \kappa_2 \phi_2^\dagger \phi_3 \phi_2^\dagger \phi_4
+ \kappa_3 \phi_1^\dagger \phi_4 \phi_1^\dagger \phi_5
+ \kappa_4 \phi_2^\dagger \phi_5 \phi_2^\dagger \phi_6
\no & &
+ \kappa_5 \phi_3^\dagger \phi_2 \phi_3^\dagger \phi_5
+ \kappa_6 \phi_4^\dagger \phi_2 \phi_4^\dagger \phi_6
+ \kappa_7 \phi_3^\dagger \phi_6 \phi_4^\dagger \phi_5
+ \kappa_7^\prime \phi_3^\dagger \phi_5 \phi_4^\dagger \phi_6
\no & &
+ \kappa_8 \phi_2^\dagger \phi_6 \phi_4^\dagger \phi_3
+ \kappa_8^\prime \phi_2^\dagger \phi_3 \phi_4^\dagger \phi_6
+ \kappa_9 \phi_2^\dagger \phi_5 \phi_3^\dagger \phi_4
+ \kappa_9^\prime \phi_2^\dagger \phi_4 \phi_3^\dagger \phi_5.
\ea

It is easy to check that with this $V_t$ VEVs
with the right powers of the suppression factor
$\epsilon \sim \left( \mew / m_H \right)^2$ are generated.
One obtains
\ba
v_3 &\approx& - \kappa_1^\ast \frac{v_1^2 v_2^\ast}{\mu_3^2},
\\
v_4 &\approx& - \kappa_2^\ast \frac{v_2^2 v_3^\ast}{\mu_4^2},
\\
v_5 &\approx&
- \kappa_3^\ast \frac{v_1^2 v_4^\ast}{\mu_5^2}
- \kappa_5^\ast \frac{v_3^2 v_2^\ast}{\mu_5^2},
\\
v_6 &\approx&
- \kappa_4^\ast \frac{v_2^2 v_5^\ast}{\mu_6^2}
- \left( \kappa_8^\ast + {\kappa^\prime_8}^\ast \right)
\frac{v_2 v_4 v_3^\ast}{\mu_6^2}.
\ea
The other terms in $V_t$ generate subdominant
(in terms of $\epsilon$)
contributions to the VEVs.

\section{Conclusions}
\label{concl}

The main point in this letter
is the observation that the VEVs of some Higgs doublets
may be suppressed by a type~II seesaw mechanism
in the same way as the VEVs of scalar gauge triplets.
We have furthermore emphasized
that this Higgs-doublet type~II seesaw mechanism
may be combined with other seesaw mechanisms of any type---I,
II,
III or even with itself in a multiply nested way.
If there are only two mass scales at our disposal,
the electroweak scale $\mew$ and a heavy scale $m_H \gg \mew$,
one may through this procedure
suppress some mass terms by a factor $\left( \mew / m_H \right)^p$,
where the power $p$ can be considerably larger than 1
as in the standard type~I seesaw case.
While the standard seesaw mechanisms
are applied to Majorana neutrinos,
the type~II seesaw mechanism for Higgs doublets,
whether in its simple or in its multiply nested form,
is able to suppress any Dirac-fermion masses
without one having to introduce
any new fermionic degrees of freedom in the theory.

Our aim was not to promote a specific type of seesaw mechanism,
rather to point out the wealth of possible scenarios.
It is also beyond the scope of this letter
to check the compatibility of each particular scenario
with the experimental data,
for instance with electroweak precision tests.
Thus,
in individual cases
the parameter space may have to be restricted or the scenario modified.

A seesaw mechanism always involves
the \textit{ad hoc} introduction of a heavy scale $m_H$.
The usual belief is that either the new physics at $m_H$
is not directly accessible by experiment because that scale is too high,
or contrived cancellation mechanisms are needed to lower $m_H$.
The main message of this letter
is that neither of the two conclusions is compelling.
As originally demonstrated in a specific case~\cite{ma1}
and generalized in this letter,
the nesting of the type~II seesaw mechanism for Higgs doublets
with other seesaw mechanisms,
or with itself,
provides a very simple method to lower $m_H$.
This method requires an extension of the scalar sector and,
therefore,
leads to new physics at the scale $m_H$.

\paragraph{Acknowledgements:}
W.G.\ and L.L.\ acknowledge support from the European Union
through the network programme MRTN-CT-2006-035505.
The work of L.L.\ was supported by the Portuguese
\textit{Funda\c c\~ao para a Ci\^encia e a Tecnologia}
through the project U777--Plurianual.
The work of B.R.\ is supported by the Croatian Ministry of Science,
Education and Sport under the contract No.~119-0982930-1016.
B.R.\ gratefully acknowledges the support of the University of Vienna
within the Human Resources Development Programme
for the selected SEE Universities
and the hospitality offered at the Faculty of Physics.

\end{document}